\documentclass[twocolumn]{emulateapj}
\usepackage[latin2]{inputenc}

\mathchardef\mhyphen="2D

\newcommand{\bssC}{\textbf{\textsf{C}}}
\newcommand{\R}{\mathcal{R}}
\newcommand{\avg}[1]{\left\langle{#1}\right\rangle}

\newcommand{\hmpc}{\,$h^{-1}$\,Mpc}
\newcommand{\hmpcnosp}{$h^{-1}$\,Mpc}
\newcommand{\ihmpc}{\,$h$\,Mpc$^{-1}$}

\newcommand{\pmsn}{P^{\rm \hspace{1pt} \mhyphen sn}}
\newcommand{\pinit}{P_{\rm init}}

\newcommand{\eg}{e.g., }
\newcommand{\logp}{{\log_{+ \hspace{-1pt}}}}
\newcommand{\plogp}{P_{\log\hspace{-1pt}+\hspace{-1pt}}}

\newcommand{\sns}{(S/N)$^2$}
\newcommand{\ncell}{n_{\rm cell}}
\newcommand{\kmax}{k_{\rm max}}
\newcommand{\knyq}{k_{\rm Nyq}}

\chardef\til=`\~

\begin{document}

\title{Rejuvenating Power Spectra II: the Gaussianized galaxy density field}

\author{Mark C. Neyrinck\altaffilmark{1}, Istv\'an Szapudi\altaffilmark{2} and Alexander S. Szalay\altaffilmark{1}}
\altaffiltext{1}{Department of Physics and Astronomy, The Johns Hopkins University, 3701 San Martin Drive, Baltimore, MD 21218, USA}
\altaffiltext{2}{Institute for Astronomy, University of Hawaii, 2680 Woodlawn Drive, Honolulu HI 96822, USA}

\begin{abstract}
We find that, even in the presence of discreteness noise, a
Gaussianizing transform (producing a more-Gaussian one-point
distribution) reduces nonlinearities in the power spectra of
cosmological matter and galaxy density fields, in many cases
drastically.  Although Gaussianization does increase the effective
shot noise, it also increases the power spectrum's fidelity to the
linear power spectrum on scales where the shot noise is negligible.
Gaussianizing also increases the Fisher information in the power
spectrum in all cases and resolutions, although the gains are smaller
in redshift space than in real space.  We also find that the gain in
cumulative Fisher information from Gaussianizing peaks at a particular
grid resolution that depends on the sampling level.

\end{abstract}


\section{Introduction}

The power spectra of fluctuations in the matter and (more observably)
galaxy fields carry important cosmological information.  On large
scales and early epochs, where the fluctuations are small and
Gaussian, this information is preserved from early epochs, each
Fourier mode having evolved linearly.  On smaller scales, when the
amplitudes of fluctuations grow to $\gtrsim 1$, the linear
approximation breaks down.  Modes of the overdensity $\delta$ become
coupled, and their evolution becomes much harder to model.
Inconveniently, we need high-order perturbation theory and numerical
simulations to model the expectation value of the power spectrum
accurately.  A more fundamental problem is that the cosmic
(co)variance in the power spectrum acquires a dominant non-Gaussian
component on surprisingly large, ``translinear'' scales
\citep{mw,szh,coorayhu}.  This has unpleasant consequences for
cosmological parameter estimation, quantified by a ``translinear
plateau'' in cumulative Fisher information content
\citep{rh05,rh06,ns06,ns07,leepen,takahashi}.

Recently, we found that performing a logarithmic transform on the
matter overdensity $\delta$, i.e.\ using $\ln(1+\delta)$ instead of
$\delta$ as a density variable, drastically reduces the nonlinearities
on translinear scales in the power spectrum \citep[][Paper I]{nss09}.
The logarithmic transform pushes the translinear plateau to scales
about 2-3 times smaller, revealing about 10 times more Fisher
information.  It also gives a power-spectrum shape intriguingly close
to the linear-theory prediction.  The density field 1-point PDF
(probability density function) is approximately lognormal
\citep{colesjones}; in fact, we found that an exact Gaussianization of
the PDF (described below) performs even better.

PDF Gaussianization in large-scale structure was first proposed by
\citet{weinberg}, although not explicitly to increase power-spectrum
information content, but to reconstruct the initial density field.
\citet{croft98} used PDF Gaussianization in processing Lyman-$\alpha$
forest data from quasar spectra, but this turns out not to be an
essential step in estimating small-scale power spectra from this
data, because radiative transfer already maps the overdensity into a
narrow range of flux \citep{croft99}.

Although PDF Gaussianization impressively recovers the shape of the
initial power spectrum, the transformation is less-successful in
reconstructing initial mode-by-mode phases and amplitudes
\citep{nssprep}.  This is because of bulk motions of matter on $\sim
10$\hmpc\ scales, and formation of the cosmic web.  For example, the
initial and final PDF-Gaussianized fields shown in Fig.\ 1 in Paper I
look by eye quite different on small scales.  Precise reconstruction
of the phases and amplitudes of translinear Fourier modes appears to
require the accurate estimation and subtraction of the Lagrangian
displacement field \citep[\eg][]{mak03,recon,lavaux08,noh}.  With a
Lagrangian reconstruction, it is obvious that the shape of the linear
power spectrum should be reconstructed on translinear scales, but the
methods are much more complicated and computationally intensive than a
simple density PDF transform.

It does make sense intuitively that PDF Gaussianization should help
the power spectrum to describe a field.  While the cosmologically
useful information in a Gaussian field is entirely in Fourier
amplitudes, the information in a non-Gaussian field is partly in phase
correlations, which are necessary to describe features such as sharp
density peaks.  Thus, flattening peaks restores information to the
Fourier amplitudes from the phases.  Phase correlations affect
higher-order statistics, not the power spectrum, so Gaussianization
can be seen as pulling information from higher-order statistics into
the power spectrum.

In the approximation that a non-Gaussian field is a non-linear
transformation of a Gaussian field, PDF Gaussianization will produce a
Gaussian field, vanquishing all higher-order correlations.
Conversely, subjecting a Gaussian field to a non-linear transformation
produces higher-order correlations \citep{Szalay88}.  In particular,
over length scales where the two-point correlation function is
positive, a monotonic transformation will generally produce a positive
four-point function, which indicates a positive non-Gaussian
contribution to the covariance through the trispectrum.  Thus it is
plausible that much of the covariance on small scales is purely from
the non-Gaussianity of the PDF.  Perhaps a related statistic to the
power spectrum of the Gaussianized field is the copula \citep{copula},
which is similarly immune to monotonic transformations on the field it
is applied to.

Despite the promising results, there remain issues to be resolved
before PDF Gaussianization can be used in practice.  In this paper, we
investigate discreteness noise.  For a logarithmic transform, the
problem becomes obvious when there are cells with zero galaxies, which
would transform to $-\infty$.  We first investigate the ideal case of
Poisson noise in the matter power spectrum, and then the galaxy power
spectrum.  We also make a start at exploring the effect of redshift
distortions.

\section{Gaussianizing transformations}
\label{sec:transforms}

There are many possible meanings of ``Gaussianizing.''  For example,
\citet{zhang10} split a density field into Gaussianized and
non-Gaussianized components based on distributions of wavelet
coefficients, and showed that the Gaussianized component of the matter
density field carries somewhat more Fisher information than the full
field.  In the present paper, by ``Gaussianization'' we mean PDF
Gaussianization, i.e.\ a function applied equally to each pixel that
reduces the higher-order moments of the one-point distribution of the
field.

We use a simple approach, first estimating the density using simple
Nearest-Grid-Point (NGP) mass assignment, and then Gaussianizing.
Perhaps some gains in information on small scales could come from
using a higher-order mass assignment scheme, or an interpolation
naturally suited to discrete data, for example the DTFE
\citep[Delaunay Tessellation Field Estimator, ][]{vdws}.
Sophisticated techniques have even been developed to estimate the
$\ln(1+\delta)$ field directly \citep[\eg][]{kitaura10,weig}.  Here we
choose NGP for its simplicity, and for the simple, constant form of
its shot noise, at least for ideal Poisson data.  More-sophisticated
techniques could perform (or inform) even better.

The two transformations we consider are ``exact'' Gaussianization,
$G(\delta)$, and a modified logarithmic transform, $\logp(\delta)$.
\citet{seo} have dealt with the problem of log-transforming zero cells
by introducing a density floor, i.e.\ adding an arbitrary small,
positive parameter to the argument of the logarithm.  This alternative
modified logarithmic transform did succeed in boosting the Fisher
information in the lensing-convergence power spectrum.  There are many
possible alternative Gaussianizing transforms.  One commonly used
transform used for producing a Gaussian distribution from a Poisson
distribution is the \citet{anscombe} transform, but this transform is
not ideal in our case.  While the separate density PDF's for each
pixel, over different realizations, should be Poisson in our case, the
global pixel density PDF will generally not be Poisson.

Our first transformation, $G(\delta)$, is the density one would expect
from an exactly Gaussian PDF with the same ranking of cell densities
as $\delta$.  Explicitly,
\begin{equation}
  G(\delta) = \sqrt{2}\sigma\,{\rm erf}^{-1}\left(2f_{<\delta}-1+1/N\right),
  \label{eqn:g}
\end{equation}
where $f_{<\delta}$ is the fraction of cells less-dense than $\delta$
in the density field, $\sigma$ is the standard deviation of the
Gaussian that $\delta$ is mapped onto, and $N$ is the number of cells.
If there are multiple cells with the same $\delta$, as usually occurs
in Poisson-sampled $\delta$'s, then there will be some range of
$G(\delta)$ that is mapped to cells with the same $\delta$.  In this
case, the actual $G(\delta)$ that we assign to these cells is an
average of $G(\delta)$ over this range.

On the other hand, a drawback of $G(\delta)$ is that it is globally
defined, nontrivially depending on the entire $\delta$ field.  Also,
the implicitly defined $G^{-1}$ function need not be well-behaved,
complicating attempts at predicting statistics of $G(\delta)$
analytically.  So, we also investigate a modified logarithmic transform,
which only depends globally on $\delta$ through the mean density.  We
define

\begin{equation}
\logp(\delta) = \left\{
\begin{array}{lc}
  \ln(1+\delta),
  & \delta > 0 \\
  \delta, & \tt{ otherwise}
\end{array}
\right . .
\label{eqn:log+}
\end{equation}

\section{Poisson-sampled matter density fields}

First we investigate the simple case of exact Poisson discreteness
noise, in the matter field investigated in Paper I.  We Poisson-sample
the density field of the Millennium Simulation \citep[MS,\ ][]{mill}
on a publicly available $256^3$ density grid at $z=0$, at various
sampling levels, from $\ncell = 1/64$ to $\ncell=64$ particles per
(2\hmpc)$^3$ cell.  (The full sampling level of the MS is
$\ncell\approx 600$.)  To Poisson-sample, we simply set the number of
particles in a cell to a random Poisson number of mean equal to the
full-sampling density.

\subsection{Effects on the mean}
It is well-known \citep[\eg ][]{peebles} that particle discreteness
produces a white-noise $1/n$ shot noise in the power spectrum, where
$n$ is the number density of particles.

\begin{figure}
  \begin{center}
    \includegraphics[scale=0.43]{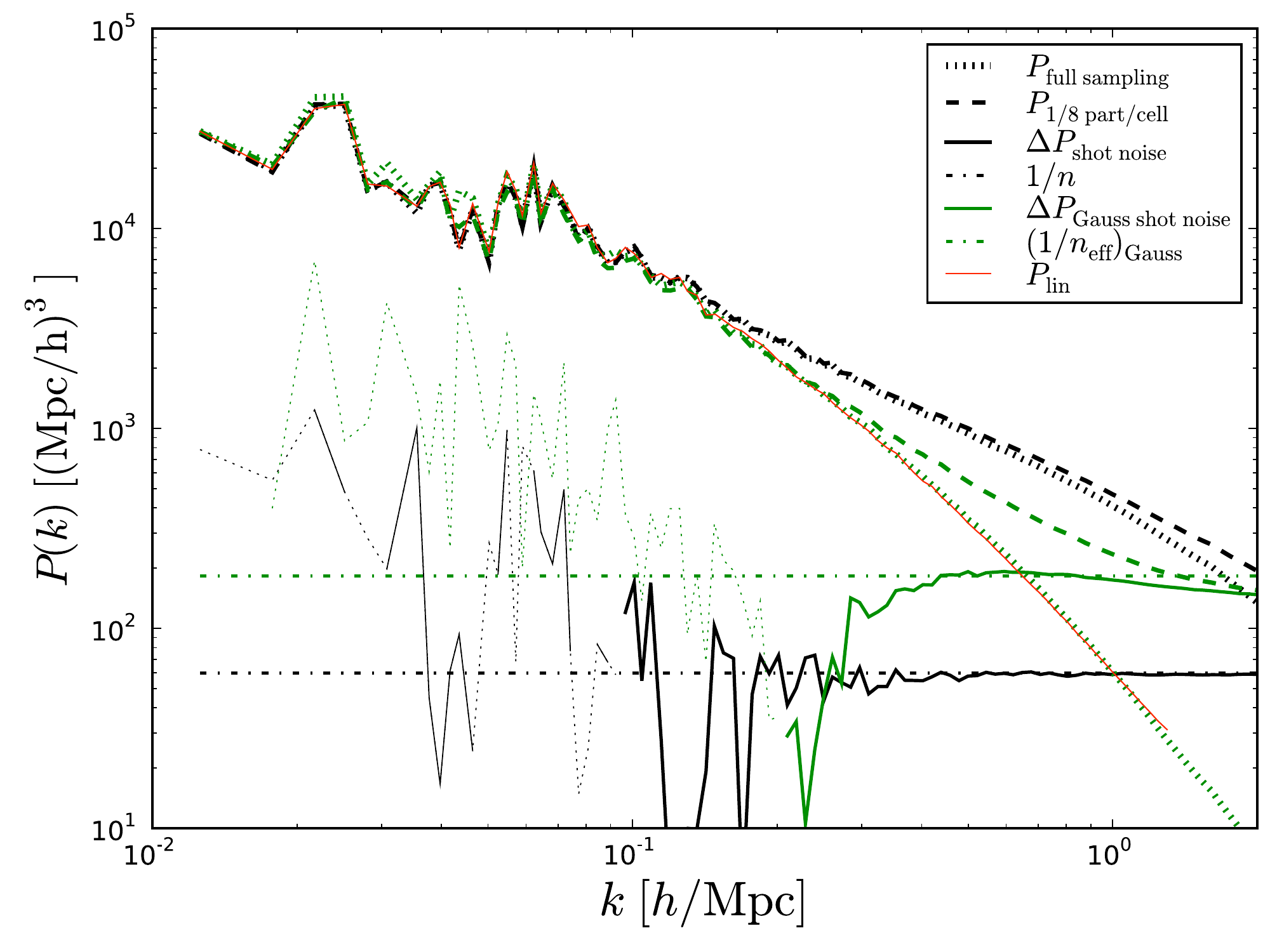}
  \end{center}  
  \caption{Poisson shot noise in the power spectra of $\delta$ (black)
    and the Gaussianized $G(\delta)$ (green), using the MS matter
    density field.  Dashed and dotted curves show $P_\delta$ and $P_G$
    at $\ncell=1/8$ and full sampling, respectively; solid black and
    green curves show their differences.  At low $k$, the absolute
    values of these differences are shown with light dotted lines.
    The solid red curve is the initial-conditions (linear) power
    spectrum, multiplied by a factor to line up with $P_{\rm \delta}$
    in the lowest-$k$ bin.  The dash-dotted line is the shot-noise
    estimate in Eq.\ (\ref{eqn:neff}).  (Some curves described here do
    not appear in the figure legend.)  }
  \label{fig:pgauss_shotnoise}
\end{figure}

Fig.\ \ref{fig:pgauss_shotnoise} shows shot noises of $\delta$ and
$G(\delta)$, estimated as the difference between power spectra of the
density fields with and without the added discreteness noise, measured
from the MS matter density field on a $256^3$ grid.  When $P_G$ and
$\plogp$ (the power spectra of $G(\delta)$ and $\logp(\delta)$) are
plotted in this paper, we multiply them by constants to line them up
with $P_\delta$ in the lowest-$k$ bin.  For $P_G$, this is equivalent
to setting the $\sigma$ used in Eq.\ (\ref{eqn:g}).

For $P_\delta$, as expected, the simple $1/n$ estimate works quite
well.  The shot noise in $P_G$, on the other hand, carries some slight
scale dependence, and is generally greater than the $P_\delta$ shot
noise.  Intuitively, Gaussianization increases the shot noise because
it increases the contrast between low-density cells.

The green dot-dashed curve in Fig.\ \ref{fig:pgauss_shotnoise} shows
an estimate of this shot noise.  It was calculated from a histogram of
$G(\delta)$, using the empirical expression
\begin{equation}
  1/n_{\rm eff}= V_{\rm cell}\sum_i f(\delta_i) (\delta_{i+1}-\delta_{i}),
\label{eqn:neff}
\end{equation}
substituting $G(\delta)_i$ for $\delta_i$.  Here, $V_{\rm cell}$ is
the volume of a cell, and $f(\delta_i)$ is the fraction of cells with
$\delta=\delta_i$, for density bins $i$.  

Eq.\ (\ref{eqn:neff}) is motivated by the low-density tail of the
density distribution, where the $G$ function stretches the contrast.
For example, in a density field with cells of only 0 or 1 particle,
$1/n_{\rm eff}=V_{\rm cell}(\delta_1-\delta_0)=V_{\rm cell}(1/N_{\rm
  particles}-0/N_{\rm particles})=1/n$.  When this density field is
transformed by $G$, the shot noise increases, proportionally with
$[G(\delta)_1-G(\delta)_0]/(\delta_1-\delta_0)$.  A simpler
approximation than Eq.\ (\ref{eqn:neff}) would be to use only $i=0,1$
in the sum (as in the preceding example), but we found that
Eq.\ (\ref{eqn:neff}) works a bit better.

\begin{figure}
  \begin{center}
    \includegraphics[scale=0.42]{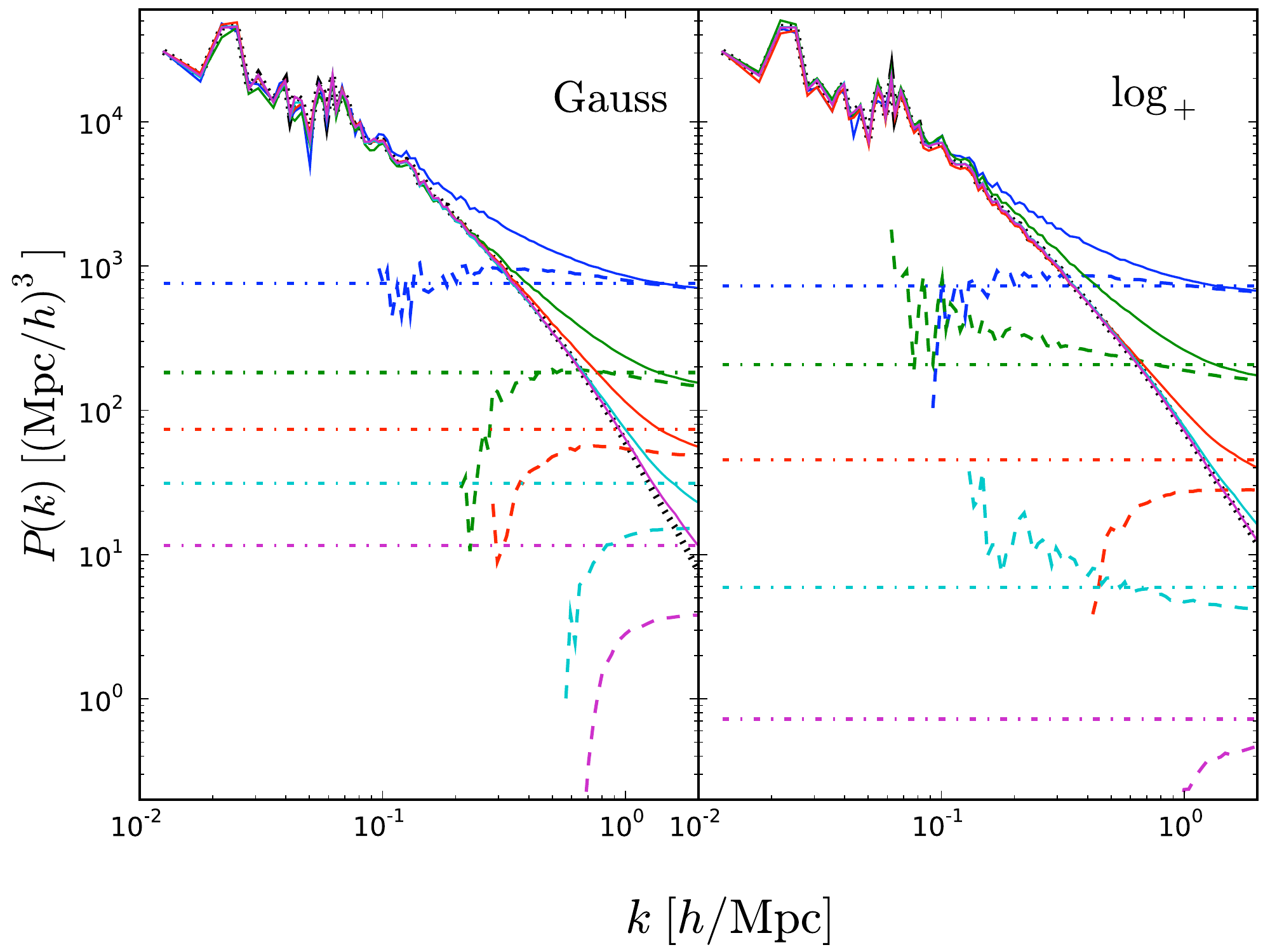}
  \end{center}
  \caption{\vspace{-2pt} Shot noise in the power spectra of
    $G(\delta)$, and $\logp(\delta)$, with varying $\ncell$, the mean
    number of particles per cell on the 256$^3$ grid.  From bottom
    (magenta) to top (blue), the number of particles per
    (2-\hmpcnosp$)^3$ cell varies from 64 to $1/64$, in multiples of
    8.  Solid curves are power spectra of the transformed
    Poisson-sampled fields, and the black dotted curves are power
    spectra of $G(\delta)$ and $\logp(\delta)$ with the full MS
    particle sampling.  Solid curves are multiplied by factors to line
    up with the dotted curves in the lowest-$k$ bin.  Dashed curves
    show the differences between the solid and dotted curves, and the
    dot-dashed lines show the shot noise estimated from
    Eq.\ (\ref{eqn:neff}).  }
  \label{fig:manyshots}
\end{figure}

Fig.\ \ref{fig:manyshots} explores the shot noise in $P_G$ and
$\plogp$ with varying sampling levels on a $256^3$ grid.  The shot
noise in $\plogp$ at high sampling and high $k$ is generally smaller
than in $P_G$; however, the shape of $\plogp$'s shot-noise curve is
less consistent than $P_G$'s.  This hints at the higher (co)variance
in $\plogp$ than in $G$, which will be discussed further in the next
subsection.  The estimate in Eq.\ (\ref{eqn:neff}) works well for low
sampling, but overestimates the shot noise if the sampling is $n_{\rm
  cell} \gtrsim 1$, especially in the $P_G$ case.

\begin{figure}
  \begin{center}
    \includegraphics[scale=0.43]{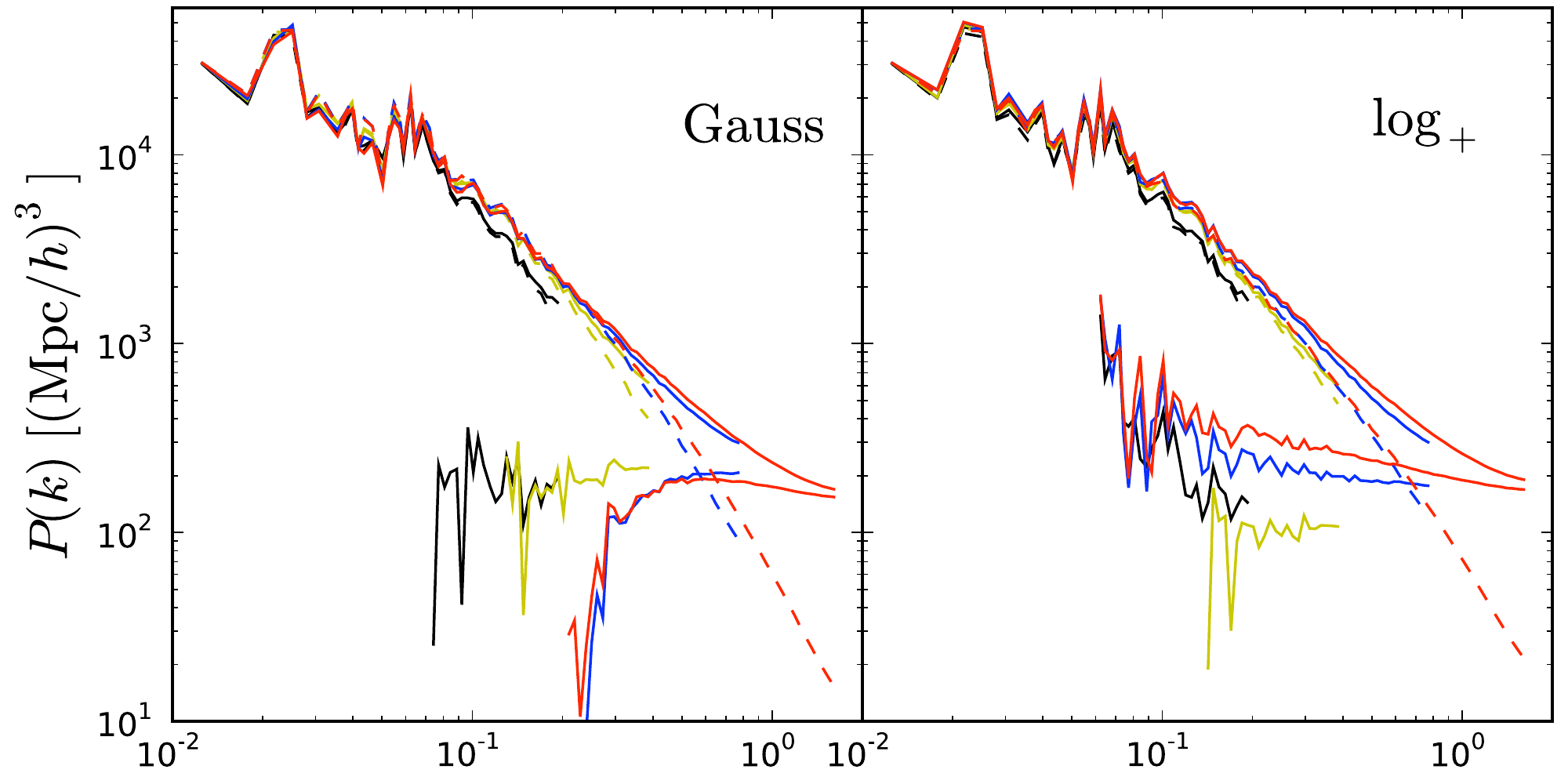}
    \includegraphics[scale=0.43]{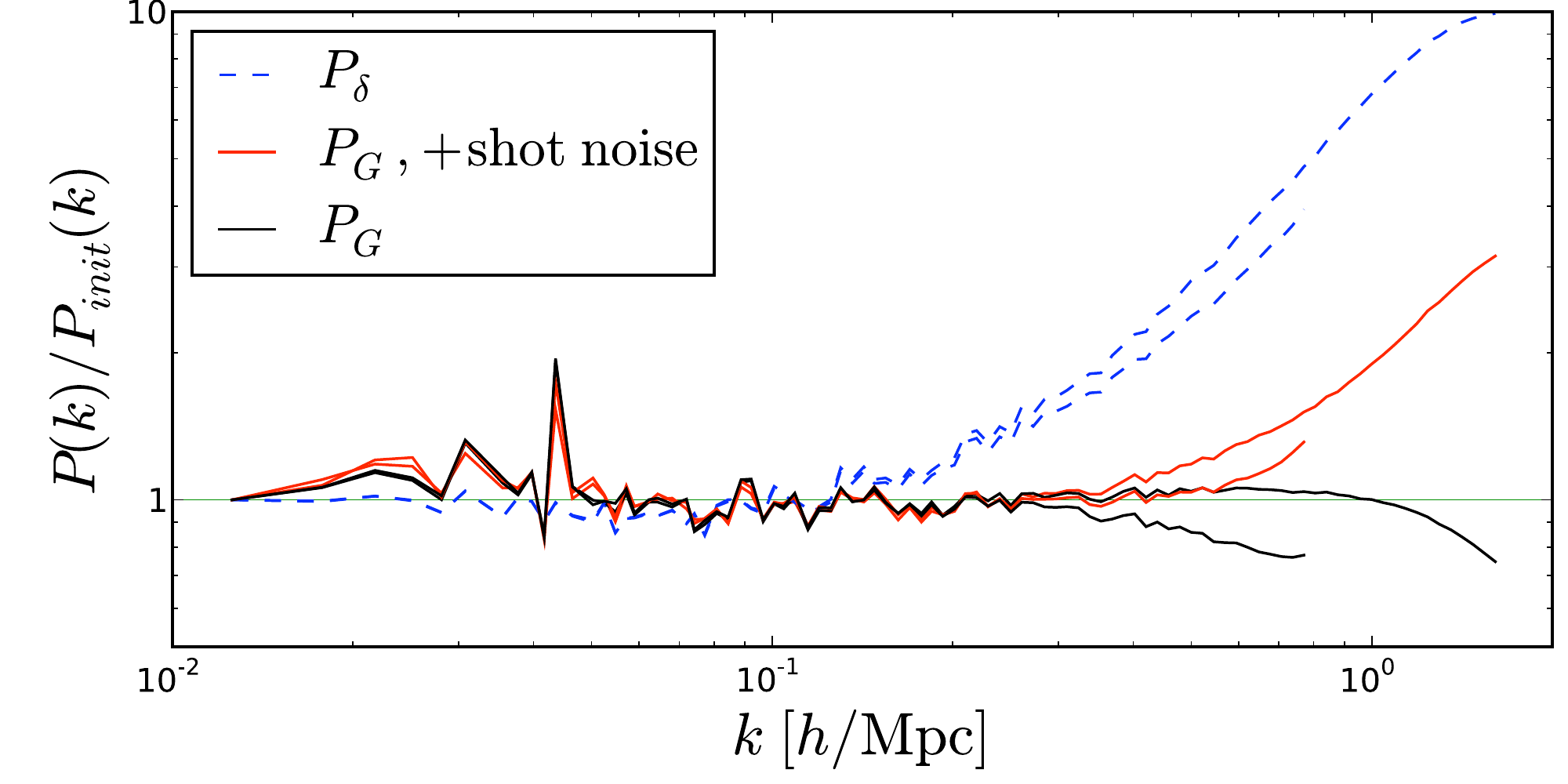}
  \end{center}
  \caption[1]{ {\it Top}.  Shot noise in the power spectra of
    $G(\delta)$, and $\logp(\delta)$, with varying grid resolution,
    for a matter density field with a fixed Poisson sampling of
    $\ncell = 1/8$ particles per (2-\hmpcnosp$)^3$ cell.  The power
    spectra are shown at grid resolutions of $32^3$ (black), $64^3$
    (yellow), $128^3$ (blue), and $256^3$ (red).  Both the power
    spectra of the Poisson-sampled field and their differences from
    the full-resolution power spectra are shown as solid curves; the
    full-resolution power spectra appear as dashed curves.  The shot
    noise is rather consistent at different resolutions, especially in
    the $G(\delta)$ case.

    {\it Bottom}. Ratios of $P_{\delta}$ and $P_G$ to the initial
    power spectrum, normalized to 1 in the lowest-$k$ bin.  $P_G$ and
    $P_\delta$ are at full sampling (as in Paper I), and ``$P_G$, + shot
    noise'' is the raw power spectrum of the density field sampled at
    $n_{\rm cell}=1$ on the 256$^3$ grid.  Ratios at two resolutions
    are shown: 128$^3$ and 256$^3$.
    \label{fig:manyres}
  }
\end{figure}

Fig.\ \ref{fig:manyres} shows how the shot noise varies with
resolution, at a fixed sampling.  Generally, especially for $P_G$, the
shot noise is rather consistent for different resolutions.  In fact,
the approximation that the shot noise is constant over different
resolutions seems to be a better approximation than the one in
Eq.\ (\ref{eqn:neff}), so we will use it when we deal with galaxies
(with no easily measurable ``no-shot-noise'' power spectrum).

The bottom panel of Fig.\ \ref{fig:manyres} shows the ``nonlinear
transfer function'' $P(k)/\pinit(k)$ for $P_G$, raw and after
subtracting shot noise, and for $P_\delta$ (after subtracting shot
noise).

\subsection{Effects on Information Content}
As in Paper I, we use a Fisher information \citep{fisher, tth}
formalism to quantify the information in the power spectrum.  The
cumulative Fisher information in the power spectrum about parameters
$\alpha$ and $\beta$ over a range of power-spectrum bin indices
$i\in\R$ is estimated as
\begin{equation}
  F_{\alpha\beta}(\R) =
  \sum_{i,j\in \R} \frac{\partial\ln \pmsn_i}{\partial\alpha}(\bssC_\R^{-1})_{ij} 
 \frac{\partial\ln \pmsn_j}{\partial\beta},
  \label{inforange}
\end{equation}
where $\bssC_\R$ is the square submatrix of $\bssC$ with both indices
ranging over $\R$.  $\bssC_\R$ is the covariance matrix of the power
spectrum in bins,
$C_{ij}=\avg{\Delta \pmsn_i \Delta \pmsn_j}/(\pmsn_i \pmsn_j) =
\avg{\Delta\ln \pmsn_i\Delta\ln \pmsn_j}$.

In Paper I, we considered the signal-to-noise ratio S/N, the
information in the power spectrum about the power spectrum itself.
(S/N)$^2$ (called simply S/N in Paper I) is the Fisher information
about a (possibly hypothetical) parameter that depends on each mode of
the power spectrum equally.  Thus, the derivative terms above were set
to unity.  For $P_\delta$, the linear-power-spectrum amplitude $A$
\citep[\eg investigated in][]{rh05,ns06} is a parameter such that
$\partial \ln \pmsn_i/\partial \ln A = 1$ on linear scales, reaching
$\approx 2$ on translinear scales.  The situation is more subtle in
the case of power spectra of nonlinearly-transformed fields, since
there is generally a large-scale bias (Paper I).  But this does not
affect parameters that depend on the power spectrum's shape.  And for
parameters that depend on the amplitude, the large-scale bias can be
constrained by measuring both $P_\delta$ and $P_G$ (or $\plogp$) in
the linear regime.  In this paper, though, we simply investigate the
S/N in $P_G$ (and $\plogp$) themselves.  This would be entirely
appropriate when comparing data to a mock catalog, for example.

Shot noise further complicates the situation.  The statistically
stable shot noise component $S_i$ of the power spectrum that appears
on small scales actually reduces the covariance in $\pmsn_i+S_i$ (the
power spectrum including shot noise), mimicking a gain in clustering
information, when really all that's being accurately measured is the
shot noise.  The correct thing to investigate is the information in
$(\pmsn_i+S_i)$ about the power spectrum without shot noise, $\pmsn_i$.  Thus
we investigate
\begin{eqnarray}
  ({\rm S/N})^2 & = & \sum_{i,j\in \R} \frac{\partial\ln(\pmsn_i+S_i)}{\partial\ln \pmsn_i}(\bssC_\R^{-1})_{ij} \frac{\partial\ln(\pmsn_j+S_j)}{\partial\ln \pmsn_j}\nonumber\\
  & = & \sum_{i,j\in \R} \frac{\pmsn_i}{\pmsn_i+S_i}(\bssC_\R^{-1})_{ij} \frac{\pmsn_j}{\pmsn_j+S_j}.
  \label{eqn:infosn}
\end{eqnarray}
For the last line, we assume that the shot noise is independent of the
clustering fluctuations.  In fact, this result is roughly what one
would get by subtracting the mean shot noise from the power spectra
before measuring the covariance matrix.  For example, if the
covariance matrix is diagonal, the Fisher-matrix entries for the power
spectrum including shot noise will be $(\bssC_R^{-1})_{ii} =
C_{ii}^{-1} = (\pmsn_i+S_i)^2/(\Delta \pmsn_i)^2$.  Plugging this into
Eq.\ (\ref{eqn:infosn}) causes the fractions to cancel, giving the
Fisher-matrix entries for the power spectrum without shot noise.

\begin{figure}
    \begin{center}
      \includegraphics[scale=0.43]{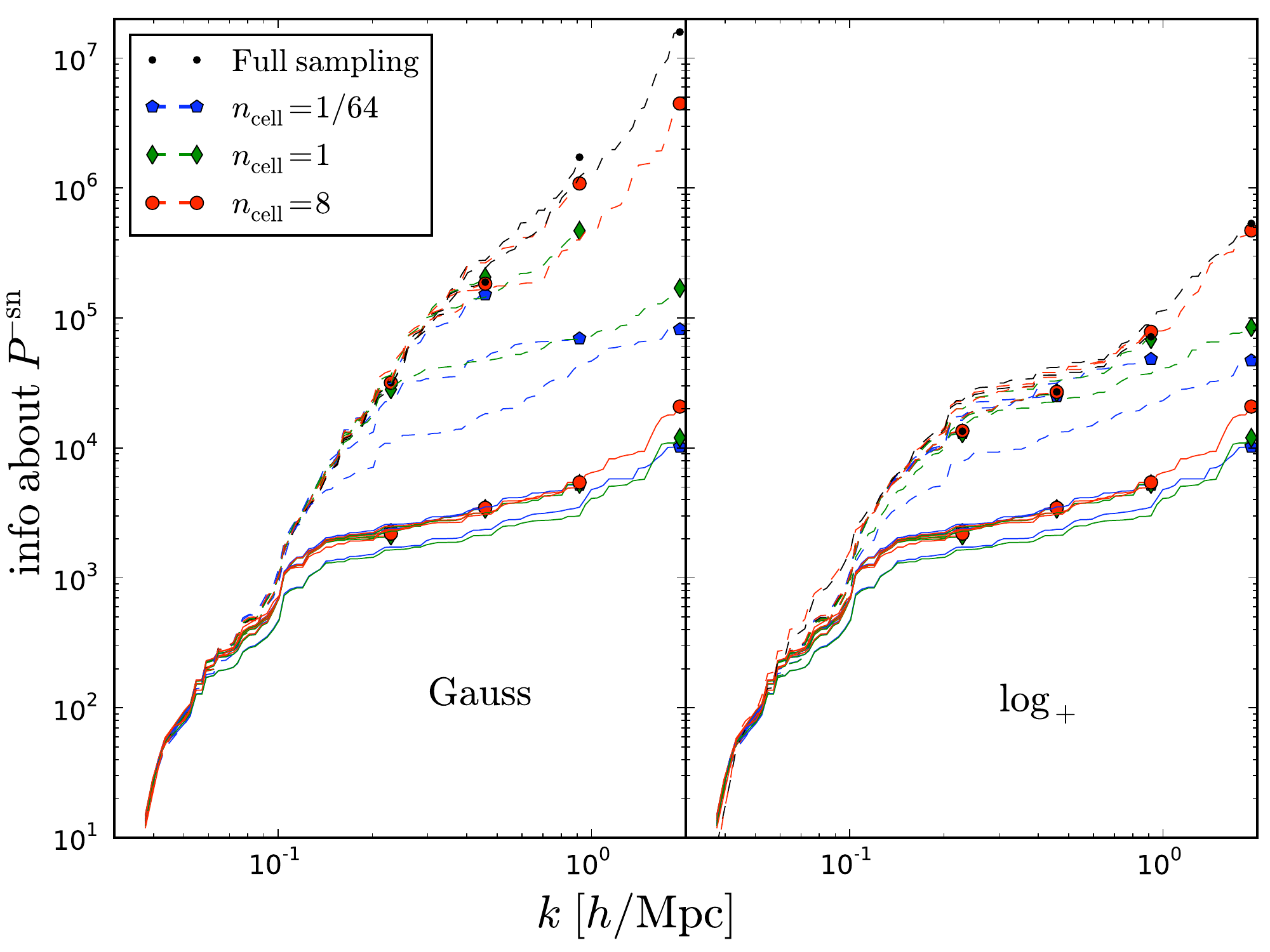}
    \end{center}
    \caption[1]{ Information \sns\ curves in the presence of
      discreteness effects, for $P_\delta$ (solid curves), and $P_G$,
      and $\plogp$ (dashed curves), using Eq.\ (\ref{eqn:infosn}).
      The plethora of curves show the results for different
      combinations of samplings and resolutions.  The three samplings
      shown are $\ncell=1/64$, 1, and 8 particles per (2\hmpcnosp)$^3$
      cell, in blue, green and red, respectively.  These $256^3$
      density grids are degraded in resolution by powers of two,
      giving $32^3$, $64^3$, and $128^3$ grids.  Information curves
      are measured for each case, and symbols are placed at their ends
      (at the Nyquist frequency).  All of these differences have
      little effect on the $P_\delta$ information, but change $P_G$,
      and $\plogp$ significantly.
      \label{fig:gausspoissinfo}    
    }
\end{figure}
  
Fig.\ \ref{fig:gausspoissinfo} shows a comparison of the \sns\ in
$P_\delta$, $P_G$, and $\plogp$, for Poisson-sampled density fields
of various resolutions and samplings of the MS.  As in Paper I,
covariance matrices are estimated from power spectra measured after
applying 248 sinusoidal weightings \citep{hrs} to the density field.
$P_G$ (dashed) out-informs $P_\delta$ (solid) in all cases, although
the gains are modest at a sampling of $\ncell=1/64$.

Interestingly, especially for $P_G$, there appears to be a resolution
at which the gains in information from Gaussianization (i.e.\ the
vertical distance between the dashed and solid curves) peak.  This is
not surprising: in the low-resolution limit, the field is already
Gaussian, so Gaussianization has no effect.  In the high-resolution
limit, even the highest peaks can only be sampled with one particle,
giving a field of only 0's and 1's, which Gaussianization will only
cause to be multiplied by a constant.  Mathematically, the information
without the $S_i/(\pmsn_i+S_i)$ fractions keeps rising, but these
fractions can cause it to turn over, producing a peak.  As we find
below in Section \ref{sec:galinfo}, the peak is generally at a
resoution a few times coarser than that where $P^{\rm
  -sn}(\knyq)\approx S(\knyq)$.  And particularly if one is interested
in the power spectrum over the range of scales just smaller than the
linear regime ($0.1 \lesssim k/$(\ihmpc) $\lesssim 0.3$), and not in
scraping information from smaller scales, it is wise to Gaussianize at
this peak resolution or coarser.

\section{Galaxy density fields}
In this section, we address the impacts of discreteness and
redshift-space distortions on the observationally relevant case of a
galaxy density field, in both real and redshift space.  Again we use
the MS, using the publicly available galaxy catalog as modelled by
\citet{delucia}.  We use three galaxy samples, with R-band
absolute-magnitude cuts $R<-17$, $R<-20$, and $R<-22$; these have
respective mean galaxy number densities of about $n=0.003$, 0.02, and
0.1 (\hmpcnosp)$^{-3}$.  We generate galaxy-density grids using NGP
density assignment from these galaxy samples.  A tiny fraction of
galaxies exactly overlapped other galaxies in position; in this case,
we keep only the brightest one.  In the redshift-space case, before
gridding the galaxies we first displace them along the $x$-axis by
$v_x/H_0$.

\subsection{Effects on the mean}
\label{sec:galmean}

\begin{figure}
  \begin{center}
    \includegraphics[scale=0.43]{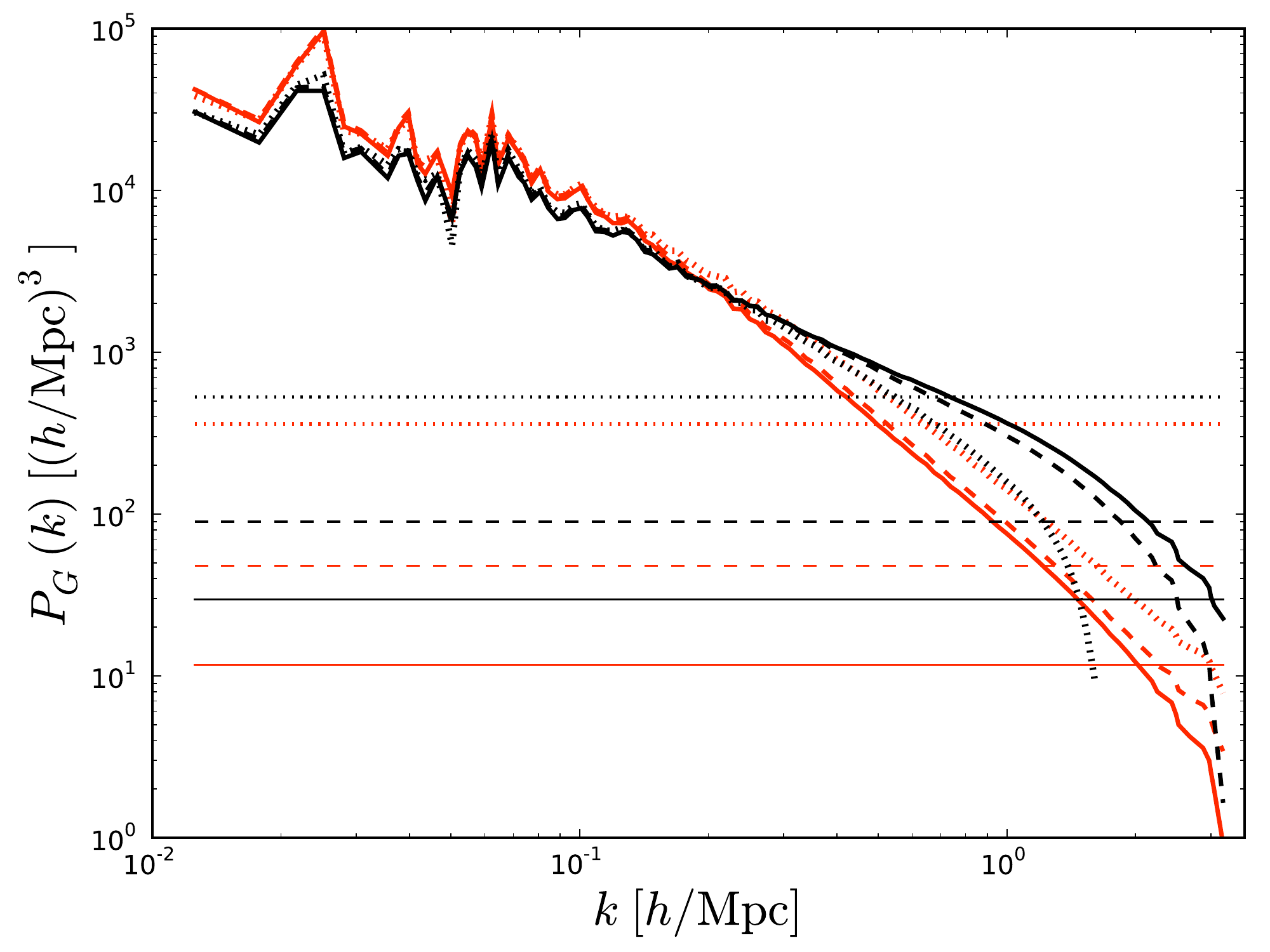}
  \end{center}  
  \caption{In bold, real- (black) and redshift-space (red) power
    spectra of Gaussianized galaxy-density fields, using MS galaxy
    samples satisfying $R<-17$ (solid), $R<-20$ (dashed), and $R<-22$
    (dotted).  Shot noise, estimated to be $P_G(\kmax)$ (see
    text), has been subtracted from them; this constant shot noise for
    each case is shown in faint lines.
  }
  \label{fig:galpg}
\end{figure}

Fig.\ \ref{fig:galpg} shows $P_G$ for galaxy density fields using the
$R<-17$, $R<-20$, and $R<-22$ samples, in real and redshift space.  In
each case, we have subtracted a shot noise, constant in $k$, and shown
with faint lines.  In this subsection we do not show $\plogp$ because
its shape in all cases is nearly identical to $P_G$.  As in the matter
power spectrum section, in showing $G(\delta)$, we set the variance of
the Gaussians onto which the $\delta$'s are mapped so that the power
spectra line up in the lowest $k$ bin.

In the matter case, the shot noise could be directly measured by
comparing to a very well-sampled matter density field, which we lack
in the galaxy case.  We estimate the shot noise using a
prescription motivated by the rough agreement of the shot noise across
different resolutions in Fig.\ \ref{fig:manyshots}.  We measure the
power spectrum on a rather high-resolution ($512^3$) grid, and assume
that for all resolutions, the shot noise is a constant with $k$, at an
amplitude of $P_G(\kmax)$.  (Here, $\kmax$ is the highest
$k$ measured in the high-resolution grid. pIn the cubic box, $k_{\rm
  max} = \sqrt{3} k_{\rm Nyquist}$, although we only plot the power
spectra to $k_{\rm Nyquist}$.)  This probably overestimates the shot
noise somewhat, but perhaps this is appropriate, at least for $P_G$,
given the slight rise in the shot-noise power at slightly lower $k$
than $\kmax$ (\eg at $k\approx 0.8$ \ihmpc\ in
Fig.\ \ref{fig:manyshots}).

\begin{figure}
  \begin{center}
    \includegraphics[scale=0.45]{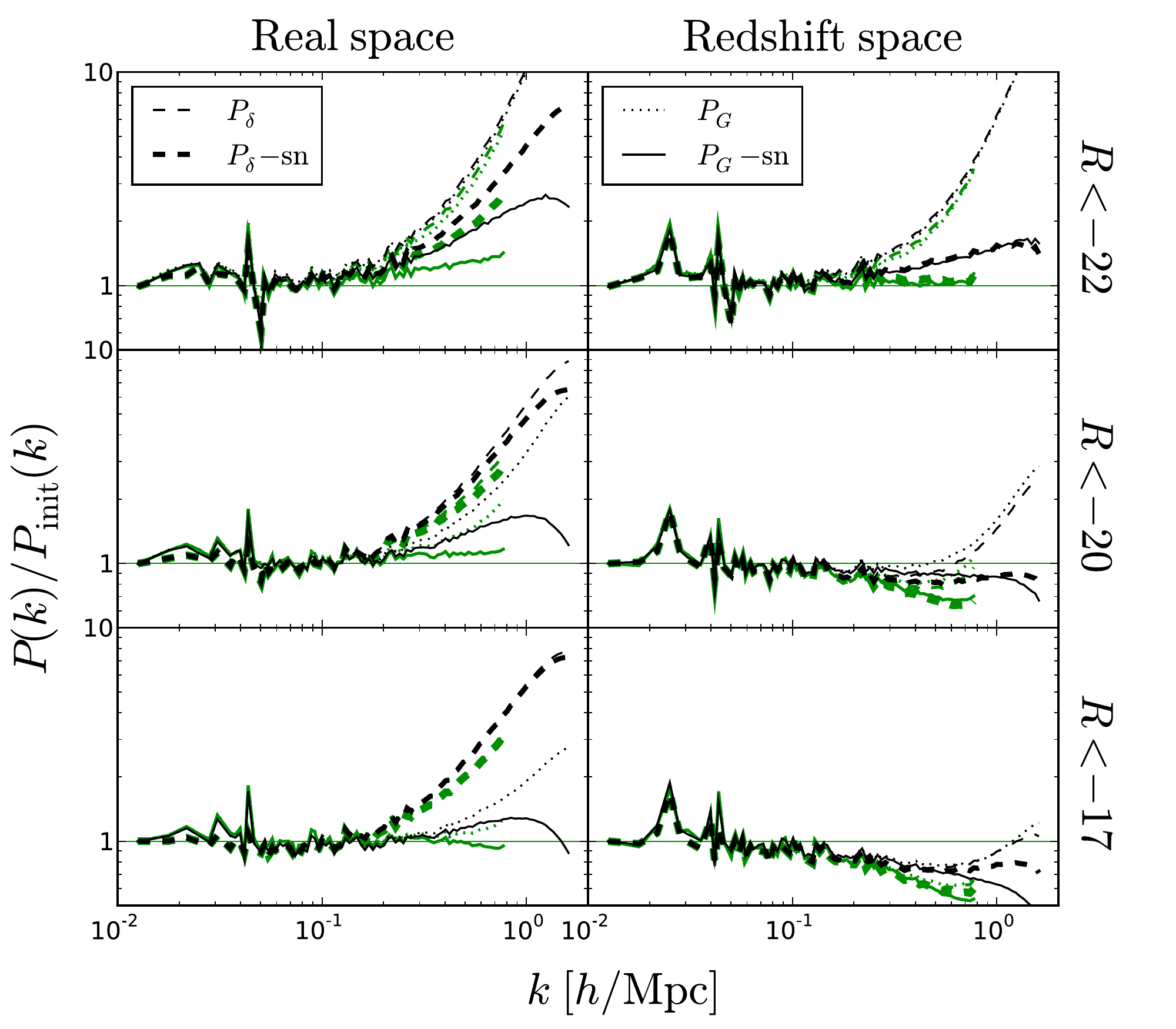}
  \end{center}  
  \caption{Nonlinear transfer functions $P_G(k)/P_{\rm linear}$ and
    $P_\delta(k)/P_{\rm linear}(k)$ for three MS galaxy samples, in
    real and redshift space.  The power spectra are measured on
    128$^3$ (green) and 256$^3$ (black) grids.  $P_G(k)/P_{\rm
      linear}$ is shown, both including (dotted), and having
    subtracted (solid), a shot-noise estimate (see text).  Thin and
    bold dashed lines show the same for $P_\delta(k)/P_{\rm
      linear}(k)$.  }
  \label{fig:galpgratio}
\end{figure}

Fig.\ \ref{fig:galpgratio} shows ratios of galaxy power spectra to the
linear power spectrum $\pinit$, in both real and redshift space.
These could be thought of as transfer functions between $\pinit$ and $z=0$ galaxy power spectra.

In real space, in the limit of low sampling (in the $R<-22$ sample),
$P_G$ and $P_\delta$ look nearly identical before shot noise is
subtracted.  This is not surprising, as $G(\delta)$ differs not much,
after removing a linear scaling, from $\delta$ in this limit.  But
after shot noise is subtracted, even at this sampling, $P_G$ seems to
track $\pinit$ a bit better than $P_\delta$.  As the sampling
increases (in the $R<-17$ sample), $P_G$ comes to track $\pinit$
significantly better, even before shot noise is subtracted.

In redshift space, the story is not as clear.  A full analysis of the
effects of redshift-space distortions on $P_G$ is beyond the scope of
this paper.  One obvious piece of analysis that is lacking is that
here we merely analyze the angle-averaged redshift-space power
spectrum.  But in general, Gaussianization modifies the shape
of galaxy power spectra less in redshift space than in real space.
This is likely because the galaxy-density PDF is already somewhat
Gaussianized because peaks are smeared by fingers of God.

\subsection{Effects on Information Content}
\label{sec:galinfo}

Our method in investigating the galaxy power spectrum Fisher
information, \sns, is essentially the same as in the matter case.  The
difference is that we do not have a meaningful measurement of the
exact shot noise, and so we estimate it as in Section
\ref{sec:galmean}.  This estimate is conservative (i.e.\ likely an
overestimate) from the point of view of information estimation, so the
information curves for $P_G$ and $\plogp$ appearing below could be
considered conservative at high $k$ (i.e.\ perhaps a slight
underestimate).  For $P_\delta$, we use the simple $1/n$ factor.

\begin{figure}
  \begin{center}
    \includegraphics[scale=0.35]{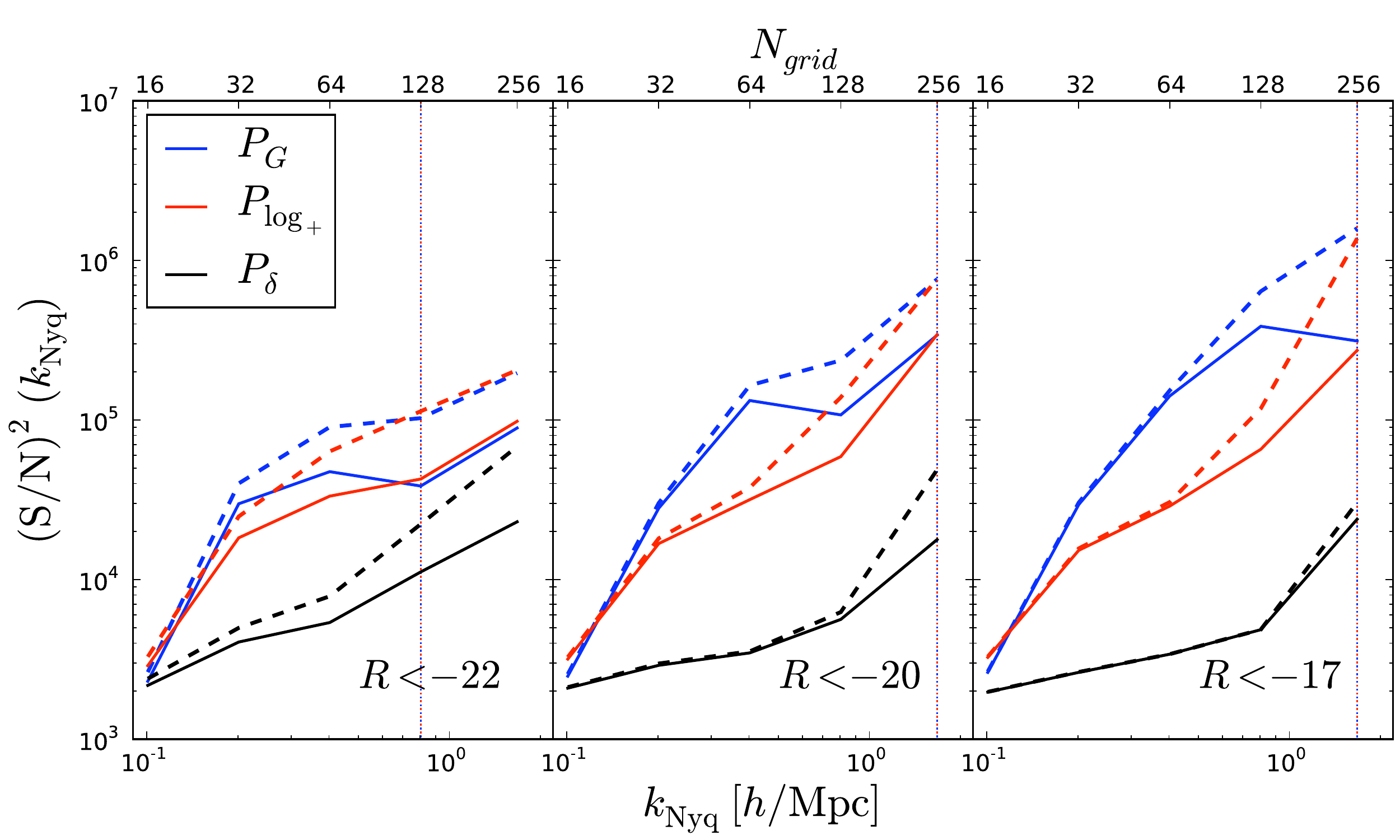}
  \end{center}  
  \caption{ Total cumulative signal-to-noise (information), up to
    $\knyq$, for the galaxy real-space $P_G$, $\plogp$, and
    $P_\delta$, computed at different resolutions and galaxy samples.
    Each point comprising the curves is analogous to a \sns curve
    endpoint-symbol in Fig.\ \ref{fig:gausspoissinfo}.  The dashed
    curves show \sns in the raw power spectra, while for the solid
    curves, the shot-noise effect has been taken into account, as in
    Eqn.\ (\ref{eqn:infosn}).  The vertical dotted lines are at the
    resolutions where, at $\knyq$, the shot-noise and clustering
    components of the power spectrum are closest.  }
  \label{fig:infogalres}
\end{figure}

Fig.\ \ref{fig:infogalres} shows \sns($\knyq$) curves for the
real-space power spectra $P_\delta$, $P_G$, and $\plogp$, for the
three MS galaxy samples, measured on grid sizes varying from $16^3$ to
$256^3$.  To reduce clutter, we do not show each cumulative \sns$(k)$
curve, but just the total cumulative \sns($\knyq$) up to the Nyquist
frequency $\knyq$.  In the matter case
(Fig.\ \ref{fig:gausspoissinfo}), these appear as symbols at
information-curve endpoints.  The dashed lines show \sns\ without
taking into account shot noise [i.e.\ without the $(\pmsn_i + S_i)/S_i$
  fractions in Eq.\ (\ref{eqn:infosn})].  The solid lines, for which
these fractions are included, are more meaningful.

Typically, as suggested in the matter case in
Fig.\ \ref{fig:gausspoissinfo}, there appears to be a peak in the gain
in \sns($\knyq$ provided by the Gaussianization transform.  The dotted
lines are at the resolution where the shot noise is most comparable to
the clustering signal; here, there is typically a trough in the
cumulative \sns. It appears that the resolution that optimizes the
gain from Gaussianizing is typically a factor of 2-4 coarser than
that. But in all cases, the Gaussianized power spectra out-inform the
standard power spectra.

\begin{figure}
  \begin{center}
    \includegraphics[scale=0.35]{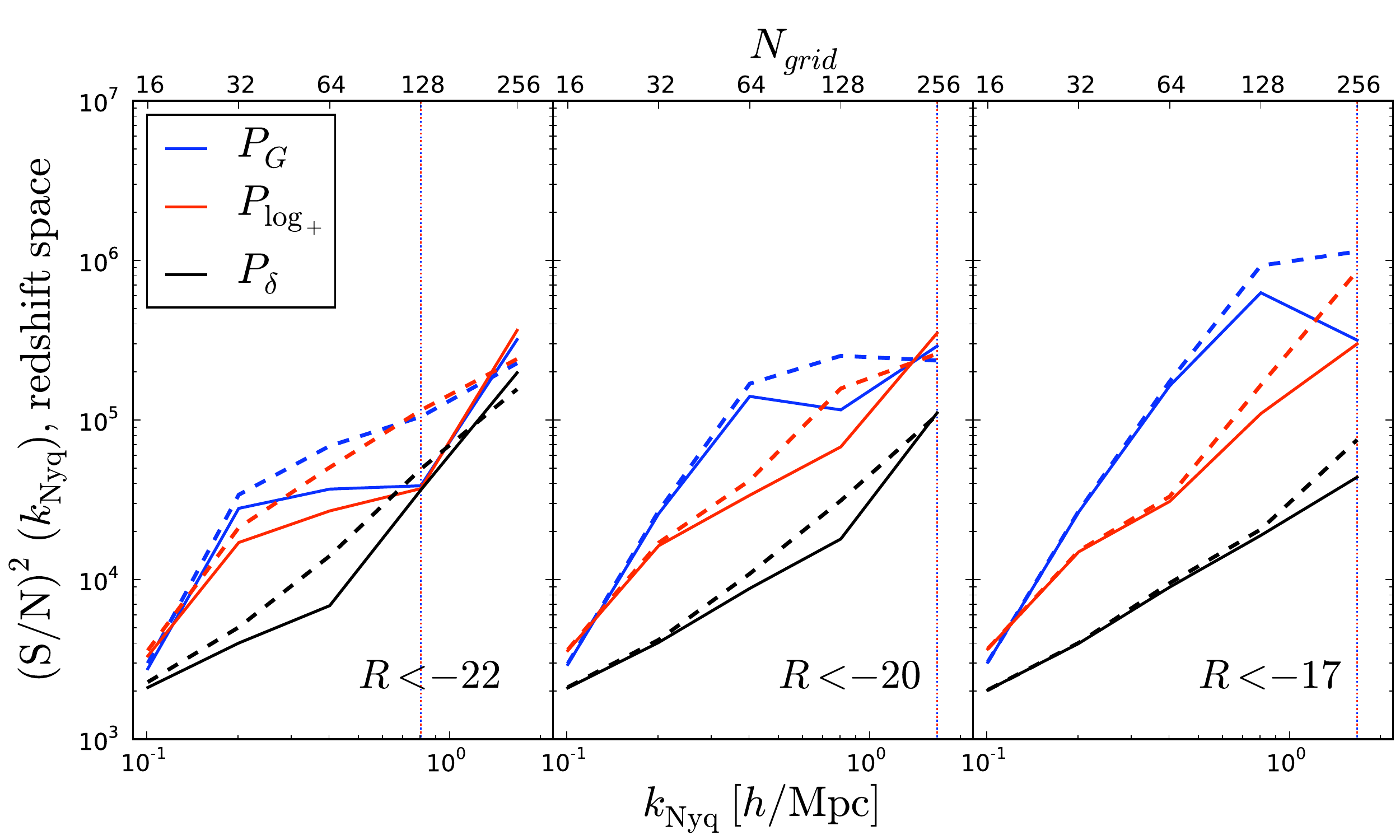}
  \end{center}  
  \caption{The redshift-space version of Fig.\ \ref{fig:infogalres}.}
  \label{fig:z_infogalres}
\end{figure}

Fig.\ \ref{fig:z_infogalres} shows the same for redshift-space power
spectra.  Here, $P_\delta$ fares better, likely because in redshift
space, fingers of God smear out density peaks and already make the PDF
of $\delta$ more Gaussian.  Still, again the power spectra of the
Gaussianized fields out-inform $P_\delta$ in all cases.

\section{Conclusion}
In this paper, we extend our previous analysis of the rejuvenating
effects that PDF Gaussianization has on the matter power spectrum
(Paper I).  We include discreteness effects, and look at the
observationally relevant case of the galaxy density field, both in
real and redshift space.  As in Paper I, we analyze the Millennium
Simulation.

We find that the conclusions of Paper I remain unchanged in the
presence of discreteness noise, as long as one is looking at scales
where the shot noise (which Gaussianization does increase somewhat) is
negligible.  In real space, Gaussianizing the galaxy and discretized
matter density fields does seem to extend the range over which their
power spectra trace the linear power spectrum, well into the nonlinear
regime, until the shot noise becomes comparable to the clustering
signal.  In redshift space, Gaussianization also reduces the
small-scale rise in the galaxy power spectrum.

Gaussianization removes or reduces the small-scale rise that one sees
in power spectra relative to the linear power spectrum.  In the
context of the halo model \citep[\eg ][]{cooraysheth}, this rise is
associated with a one-halo term.  It is perhaps not surprising that
Gaussianizing would reduce this one-halo term, since haloes are the
most non-linear, non-Gaussian structures in the Universe.  However,
the removal of this rise in the galaxy power spectrum without direct
mention of haloes also suggests that explaining galaxy bias with a
non-linear transformation of a Gaussian field \citep[\eg
][]{politzerwise,Szalay88}, which has somewhat gone out of fashion,
may be a fruitful area for further study.

In all cases, Gaussianization also improves the inherent Fisher
information, \sns, of the power spectrum.  But the degree of help it
provides depends on the resolution of the grid over which the PDF is
Gaussianized.  It seems that the grid size providing the most
cumulative added \sns\ is a factor of 2-4 coarser than the resolution
where, at the grid's Nyquist frequency, the shot noise and clustering
components of the power spectrum are of comparable magnitude.  That
is, to reap the information gains of Gaussianization on translinear
scales, one should be careful not to use grid cells that are too
small.  In redshift space, the gains in \sns\ for galaxy density
fields are somewhat smaller than in real space, if the galaxy sampling
is high enough to resolve fingers of God.  This is because fingers of
God smear high density peaks, producing an already more-Gaussian PDF.

While discreteness effects are an essential issue to investigate in
the study of power spectra of Gaussianized fields, a few issues still
remain.  With a variable survey selection function, it may be
necessary to apply a Gaussianization transform seperately in different
redshift shells.  We have made a start at analyzing the effects of
redshift distortions, but much more work can be done in this area.  It
also remains to be investigated precisely how faithfully, and to what
scales, the power spectrum of the Gaussianized matter and galaxy
density fields traces the linear power spectrum, for arbitrary
cosmologies.  Put more practically, the Fisher-matrix analysis needs
to be extended to particular cosmological parameters.  Also, our
assertion that Gaussianization pulls information from higher-point
statistics could do with further quantitative elucidation.

\acknowledgments We thank Andrew Hamilton for useful discussions, in
particular for the $\logp$ transform idea.  The Millennium Simulation
databases used in this paper and the web application providing online
access to them were constructed as part of the activities of the
German Astrophysical Virtual Observatory.  MN and AS are grateful for
support from the W.M.\ Keck and the Gordon and Betty Moore
Foundations, and IS from NASA grants NNG06GE71G and NNX10AD53G.

\bibliographystyle{hapj}
\bibliography{refs}

\end{document}